\documentclass[conference,compsoc]{IEEEtran}
\ifCLASSOPTIONcompsoc
  \usepackage[nocompress]{cite}
\else
  \usepackage{cite}
\fi
\usepackage{color}
\usepackage{graphicx}
\usepackage{grffile}
\DeclareGraphicsExtensions{.pdf,.jpeg,.png,.eps}
\usepackage{url}

\usepackage{tikz}
\usepackage{subcaption}
\usepackage{multirow}

\newcommand\Tstrut{\rule{0pt}{2.6ex}}       
\newcommand\Bstrut{\rule[-0.9ex]{0pt}{0pt}} 
\newcommand{\TBstrut}{\Tstrut\Bstrut} 

\newcommand{\eat}[1]{}
\begin{document}
%
\title{EdgeBench: A Workflow-based Benchmark for Edge Computing}

\author{\IEEEauthorblockN{Qirui Yang, Runyu Jin, Nabil Gandhi}
\IEEEauthorblockA{Arizona State University\\
Email: qyang30@asu.edu \\
rjin9@asu.edu \\
ngandhi7@asu.edu}
\and
\IEEEauthorblockN{Xiongzi Ge, Hoda Aghaei Khouzani}
\IEEEauthorblockA{NetApp\\
Email: Xiongzi.Ge@netapp.com \\
Hoda.AghaeiKhouzani@netapp.com }
\and
\IEEEauthorblockN{Ming Zhao}
\IEEEauthorblockA{Arizona State University\\
Email: mingzhao@asu.edu}}


%


\maketitle

\begin{abstract}
Edge computing has been developed to utilize multiple tiers of resources for privacy, cost and Quality of Service (QoS) reasons. Edge workloads have the characteristics of data-driven and latency-sensitive. Because of this, edge systems have developed to be both heterogeneous and distributed. The unique characteristics of edge workloads and edge systems have motivated EdgeBench, a workflow-based benchmark aims to provide the ability to explore the full design space of edge workloads and edge systems. EdgeBench is both customizable and representative. It allows users to customize the workflow logic of edge workloads, the data storage backends, and the distribution of the individual workflow stages to different computing tiers. To illustrate the usability of EdgeBench, we also implements two representative edge workflows, a video analytics workflow and an IoT hub workflow that represents two distinct but common edge workloads. Both workflows are evaluated using the workflow-level and function-level metrics reported by EdgeBench to illustrate both the performance bottlenecks of the edge systems and the edge workloads. 
\end{abstract}


%
\IEEEpeerreviewmaketitle

\section{Introduction}
\label{sec:intro}
\vspace{-5pt}
The proliferation of the IoT devices and the rapid growth of the IoT data have called forth the edge computing. In early settings, edge devices produced data and sent it to cloud for computation and storage. The high costs of cloud services and the network latency caused by data transportation outweigh the high performance cloud can provide. Edge computing emerged to solve the problem.
The computation and storage capability are brought from the cloud to the edge data centers and IoT devices for privacy, cost and Quality of Service (QoS) reasons. This has resulted in the three tiers of edge computing: IoT tier (IoT devices), edge tier (edge servers) and the cloud tier (cloud data centers). 

Contrary to the rapid growth in applying IoT devices and edge systems, there is a lack of suitable benchmark sets that capture the unique characteristics of edge workloads and edge systems. 
First, edge workloads are data-driven. The ubiquitous edge devices can generate 400 Zettabytes of data per year~\cite{biookaghazadeh2018fpgas}. This vast amount of data has raised challenges to both data transportation and data storage. Based on the various forms and sizes of the edge data, data is stored differently. For small sensor data, lightweight and scalable messaging queues are mostly used for fast and reliable data storage. For large video files, persistent storage is more preferred such as object storage or file systems. Second, many edge workloads are latency sensitive. Although the cloud tier has the strongest computation power, the network latency fails to meet the edge workloads latency requirements. These edge applications prioritize IoT tier over cloud tier for low latency. Other edge applications also switch to utilize IoT tier and edge tier more often and only offload some offline and computation expensive workloads to the cloud tier. 

To satisfy the unique requirements of edge workloads, edge systems are designed both heterogeneous and distributed where heterogeneous computing resources from different physical locations work coordinately to deliver services to end users. Different computing models such as microservices, Function-as-a-Service (FaaS) have been developed to enable computing across the heterogeneous and distributed resources on the devices, edge and cloud. These computing models have also become an essential part of edge systems. Edge benchmarks should consider the evaluation of both the edge workloads and edge systems.

Existing benchmarks~\cite{gan2019open, mcchesney2019defog, wang2018cavbench, sridhar2017evaluating} fail to capture the unique characteristics of edge workloads and edge systems. They predefine the edge workloads, the data storage backend and where each workflow stage executes. The benchmarks are not generalized to capture different edge workloads and users cannot customize the workloads for different edge scenarios. 

The unique characteristics of edge workloads and edge systems have motivated us to introduce EdgeBench, a workflow-based benchmark. EdgeBench allows users to self-define an edge workflow by providing the workflow logic. The workflows can be a sequence of pipeline, a set of cron jobs with data communication among the jobs, or some independent functions accessing the same set of data. EdgeBench orchestrates different workflow stages into workflows following the workflow logic. To capture the data-driven characteristic of edge workloads, EdgeBench offers users the flexibility to define different data storage backends for each individual function. Given the edge systems are designed to be heterogeneous and distributed, EdgeBench supports different workflow stages to be distributed across the IoT tier, the edge tier and the cloud tier and offers users the freedom to explore the tradeoff between cost and latency. Users can define where to run each function and EdgeBench automates the cross-tier execution. EdgeBench provides both function-level and workflow-level performance metrics to help users better understand their workflow and the edge system.

We implemented two representative workflows using EdgeBench: a video analytics workflow and a IoT hub workflow. The video analytics workflow forms a pipeline of continuous video streams generation, motion detection, face detection, and face recognition. The whole pipeline involves four stages and uses Minio~\cite{minio} at the IoT and edge tier and AWS S3~\cite{s3} at the cloud tier for storage of input, output and intermediate data. The IoT hub workflow mimics the sensor behaviors of hundreds of thousands IoT devices. Each IoT device periodically sends sensor data to the database and users can conduct SQL queries and data prediction using machine learning algorithms on the sensor data. This workflow involves four jobs: sensor data generator, LSTM training, LSTM prediction, and query and uses InfluxDB~\cite{influxdb} at the IoT and edge tier and AWS Timestream database~\cite{timestream} at the cloud tier for sensor data storage. It uses VerneMQ~\cite{vmq} MQTT~\cite{mqtt} broker for sensor data delivery.

We prototyped EdgeBench on OpenFaaS~\cite{openfaas}, a FaaS-based computing model. We use the two edge workflows to study the usability of EdgeBench. First, we show EdgeBench can be utilized to discover both the edge systems and the edge workloads performance bottlenecks. By providing metrics of CPU, I/O (storage and network), and memory resources, we can quantitatively determine which function is the performance bottleneck. Also, users can explore different storage backends with the resource metrics and decide which storage backend is the best fit for the edge workflow. Our evaluation results show that in the video analytics workflow, face recognition is the most computation intensive stage. In the IoT hub workflow, LSTM training is the most computation intensive job and can largely affect other jobs' performance. We also show that existing edge systems distribute loads unevenly across resources and some heavily utilized data servers can become the bottleneck. Second, we evaluate the system's performance when running functions across different tiers with various workloads. The application's end-to-end runtime latency and each function's runtime latency are reported. Users can analyze the performance results and decide the appropriate function orchestration across different tiers. Our evaluation results show that the video analytics workflow should be applied across the three tiers to utilize the benefits from different resources. For the IoT hub workflow, users should distribute the computation intensive job LSTM training on the cloud tier to improve its performance and reduce its interference with other jobs.

As edge computing and edge applications continue to evolve, it is essential for edge benchmarks to also evolve with them, to ensure that their prevalence does not come at a performance and/or efficiency loss. EdgeBench is developed to help academic and industrial researchers and developers in application development, system resource management and cross-tier acceleration. It will be open-sourced to encourage more research in this emerging field.

In summary, the contribution of EdgeBench are as follows: 
\begin{itemize}
\item Customization of workflow logics, data storage backends and execution tiers on the edge workloads.
\item Identification and evaluation of two representative edge workloads
\item Report of function-level performance metrics to explore the performance of both edge workloads and edge systems

\end{itemize}

\section{Background and Motivation}
\label{sec:back}
\vspace{-5pt}
As edge computing continues to evolve, application and system developers and researchers face many design choices. EdgeBench is designed to help developers to easily explore different design choices and encourage the best design for both edge applications and edge systems. 

First, EdgeBench considers the design choice of workflow logics. The diverse edge workloads consist of various workflow logics. For example, in video analytics workflow, one function gets executed after another and the last stage's output serves as the next stage's input. This pipelined workflow is also commonly seen in edge scenarios such as autonomous driving, personal assistant. Some other edge workloads consist of a bunch of cron jobs. For example, in a truck company management system, sensors periodically generate data and store them in a database. Other query jobs, data analytics jobs periodically get triggered and process the data. This cron job workflow is largely applied in IoT environments, such as smart agriculture or smart buildings. 

Second, EdgeBench considers the design choice of data storage backends. One notable characteristic of edge computing is the vast amount and various types of data edge devices can generate~\cite{yang2019smartdedup}. The data storage remain a significant design choice for I/O performance. For most IoT applications, sensor data generated by IoT devices is small and episodic. Lightweight message queues such as MQTT are suitable for sensor data storage. On the other hand, for video streaming applications, large video files are generated and transferred across different functions. In these cases, persistent storage such as object storage or filesystems are preferred to be used for storing and transferring data. The multi-tier development of edge computing has further complicated this design choice. Accessing data across tiers adds the network latency to the total execution time.   

Third, EdgeBench considers the design choice of function distribution on different tiers. In one workflow, some functions have strict latency requirements and should be distributed near the devices. For computation and storage expensive functions, they should be distributed to utilize the vast resources of the cloud. Also, some functions may be allocated to utilize the heterogeneous resource on the computing tier. The mix and match of functions with different computing tiers can largely affect the performance. 

In the following section, we will detail the design of EdgeBench and present the two edge workflows developed using EdgeBench.   

\section{EdgeBench Design}
\label{sec:design}
\vspace{-5pt}
\subsection{Design Overview}
\vspace{-5pt}
Overall the design of EdgeBench is based on the following key principles:

\begin{itemize}
\item \textbf{Representativeness:} The diverse edge workloads have distinct data input and output, multiple computation stages, and various latency requirements, so EdgeBench should be as representative as possible to capture all kinds of characteristics of edge workloads. 

\item \textbf{Customizability:} Both edge workloads and edge systems contain customizable components for different edge scenarios such as workflow logics, distribution of computation stages, and the storage backends for data of each stage. EdgeBench should allow the customization of all of these and the full exploration of the design space. 
\end{itemize}

While following these general principles, EdgeBench also provides a set of predefined useful workflow logics, data storage backends and two representative edge workloads. 

EdgeBench can be implemented using different computing models (monolithic applications, microservices, or Function-as-a-Service (FaaS)). EdgeBench chooses function-based abstraction, which allows users to exploit the convenience of deploying computing kernels as functions and deploying the functions on heterogeneous and distributed resources. For long-running storage services in edge workloads, EdgeBench uses microservices to allow easy-deployment and customizability of different storage backends. 

According to the first Design Principle, Representativeness, EdgeBench implements two distinct workflows, video analytics workflow and IoT hub workflow, that represent a variety of edge workloads. The video analytics workflow generates continuous large video streams and conducts intensive machine learning and computer vision algorithms on the video streams. Both the CPU usage and I/O throughput of this workload are substantial. This workflow covers common edge scenarios such as autonomous driving, drone swarm, surveillance system, and security system. The IoT hub workflow, on the other hand, generates quite different workloads from the video analytics workflow. It involves hundreds of thousands of scattered wimpy sensors which are discretely generating small and lightweight IoT data. Although each piece of data is small, the total amount is vast. This workload is more representative of edge scenarios that are related to IoT environment such as home automation, smart agriculture, smart industry, smart building, and smart city. Both workloads involve different forms and sizes of data and multiple computing stages which are representative of edge workloads.

According to the second Design Principle, Customizability, EdgeBench allows the customization of workflows, data storage backends, and the distribution of workflow stages. Users can define all of them using a per-function function template (a yaml file). EdgeBench parses the function template to orchestrate individual functions into user-defined workflows. EdgeBench provides a uniform interface for users to implement different storage backends. To distribute workflow stages among different computing tiers, users input the desired computing tier of a function and EdgeBench automates the cross-tier execution. Altogether, these allow users to fully explore the design choices of different edge workloads and edge systems. 

The rest of this section explains the various customizable components of EdgeBench, the two representative edge workloads and the reported metrics by EdgeBench. The core design of EdgeBench is applicable to all kinds of FaaS computing models. We use our prototypes for OpenFaaS, a popular open-source serverless platform, to explain EdgeBench.

\subsection{Customizable Components of EdgeBench}
\vspace{-5pt}

\begin{figure}[t]
    
	\centering
	\includegraphics[width=0.8\columnwidth]{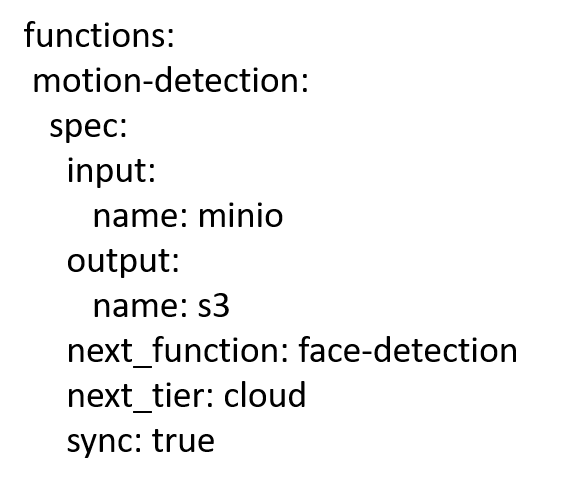}
	\caption{Function Template Snippet}
	\label{fig:code}
	\centering
\end{figure}

\begin{figure}[t]
    
	\centering
	\includegraphics[width=0.95\columnwidth]{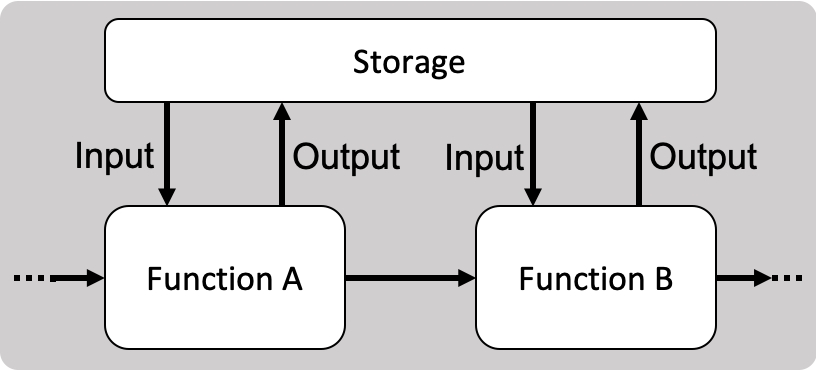}
	\caption{EdgeBench Workflow}
	\label{fig:workflow}
	\centering
\end{figure}

\begin{figure*}[t]
	\begin{subfigure}[b]{0.224\textwidth}
		\includegraphics[width=\textwidth]{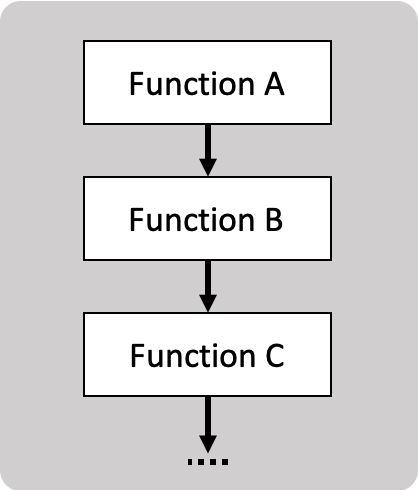}
		\caption{Pipeline}
		\label{fig:pipeline}
	\end{subfigure}
	\hspace{10pt}
	\begin{subfigure}[b]{0.224\textwidth}
		\includegraphics[width=\textwidth]{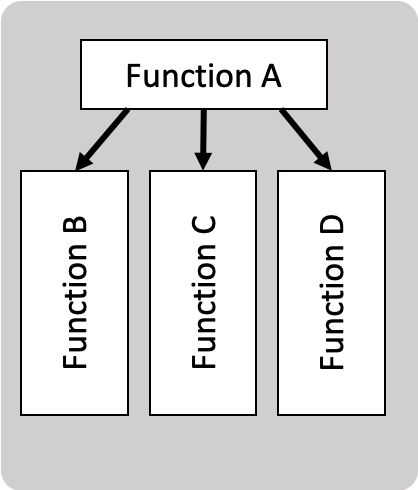}
		\caption{One-to-many}
		\label{fig:One-to-many}
	\end{subfigure}
	\hspace{10pt}
	\begin{subfigure}[b]{0.224\textwidth}
		\includegraphics[width=\textwidth]{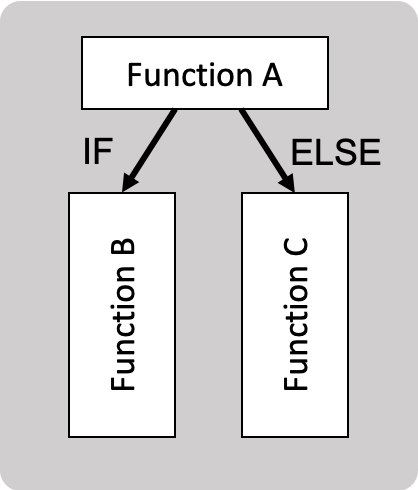}
		\caption{Branching}
		\label{fig:Branching}
	\end{subfigure}
	\hspace{10pt}
	\begin{subfigure}[b]{0.224\textwidth}
		\includegraphics[width=\textwidth]{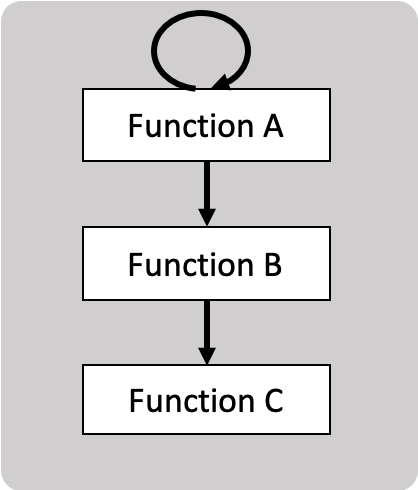}
		\caption{Cron Workflow}
		\label{fig:Cron}
	\end{subfigure}
	\caption{EdgeBench Defined Workflow Logics}
	\label{fig:logics}
\end{figure*}

\smallskip
\noindent
\textbf{Workflow.}
In EdgeBench, users define a workflow using a function template (a yaml file). There are several entities in the template that allow users to define the workflow logic, data storage backends and the execution tier. The entities are: input, output, next\_function, next\_tier, sync. Input defines the storage name of the input data for this function. Output defines the storage name of the output data for this function. Next\_function is the function name of the next stage. Next\_tier is the desired execution tier of the next function. This specifies the computing tier to execute the next function. Sync defines how the function is invoked, synchronously or asynchronously. Figure~\ref{fig:code} shows a function template snippet that uses Minio as the input source and AWS S3 as the output destination. When the function is invoked, EdgeBench fetches the input data from the user-specified input source and feeds it to the function. After the function finishes, the output data is returned to EdgeBench which saves the data to the corresponding output destination. The next function is called using the RESTful API provided at the next\_tier. The whole flow is implemented as a wrapper of the function handler and users can swap the function handlers for different functions. Figure~\ref{fig:workflow} illustrates the EdgeBench workflow.

Based on this basic workflow, we generalize four useful workflow logics that we found are popular in edge workflows: pipeline, one-to-many, branching, and cron workflow (Figure~\ref{fig:logics}).

The pipeline workflow (Figure~\ref{fig:pipeline}) forms a pipeline of functions. Each function's output is specified the same as the next function's input. EdgeBench executes each function in order and passes the output from one to the input of the other. In the pipeline, the last function leaves the next\_function and next\_tier as empty entities to identify itself as the last function.

The one-to-many workflow (Figure~\ref{fig:One-to-many}) allows many functions to be executed after one function finishes. In this case, users can specify next\_function 1 to next\_function N, in the function template to represent all the function executions following the current function. All the next\_functions should specify the same input as the current function's output to correctly accept the input data. Each next\_function can specify their own next\_function and the whole workflow forms like a tree structure. 

The branching workflow (Figure~\ref{fig:Branching}) works like if-else conditional branching. The first function generates outputs based on different conditions. Following different outputs, the next\_function will be different. Although branching workflow looks similar as one-to-many workflow, only one of the next\_functions satisfies the condition and gets invoked. In the function template, in addition to specifying next\_function 1 to next\_function N, users need to specify output 1 to output N. EdgeBench matches the data name of the function output to the data name specified in the output of the function template to decide which branch is executed next. Output and next\_function that come with the same number are bundled in one branch. 

The cron workflow (Figure~\ref{fig:Cron}) sets the whole workflow as a cron job. Users specify the cron entity with the keyword cron in the function template to illustrate how often the workflow is invoked. The time duration can be set from seconds to hours. The cron workflow can be either an end-to-end workflow involving many functions or a single function.

\smallskip
\noindent
\textbf{Data Storage Backends.}
Statelessness issue still exists with their ability to cache data ephemerally, though. In most edge workloads, functions need to access data on external storage.
Besides the above workflow logic, EdgeBench provides data storage backends for the input and output data of each function. 
Users can use their own storage backends by implementing the data storage interface provided by EdgeBench. The interface includes load() and store() functions to be used for getting input data and storing output data. EdgeBench implements three popular storage backends using the interface for input and output data: Minio~\cite{minio}, AWS S3~\cite{s3}, and Apache Kafka~\cite{kafka}. Users can use them directly by specifying the name of the storage backend as minio, s3, and kafka in the function template.

\smallskip
\noindent
\textbf{Distribution of Functions.}
To execute functions at different tiers, users first deploy the function on the desired tier and then provide the next\_function and the next\_tier. EdgeBench invokes the next\_function on the specific computing tier through the RESTful API and automates the execution across the tiers. This enables each function to be executed on different tiers.

\subsection{Two Representative Workflows}
\vspace{-5pt}

\begin{figure}[t]
    
	\centering
	\includegraphics[width=0.95\columnwidth]{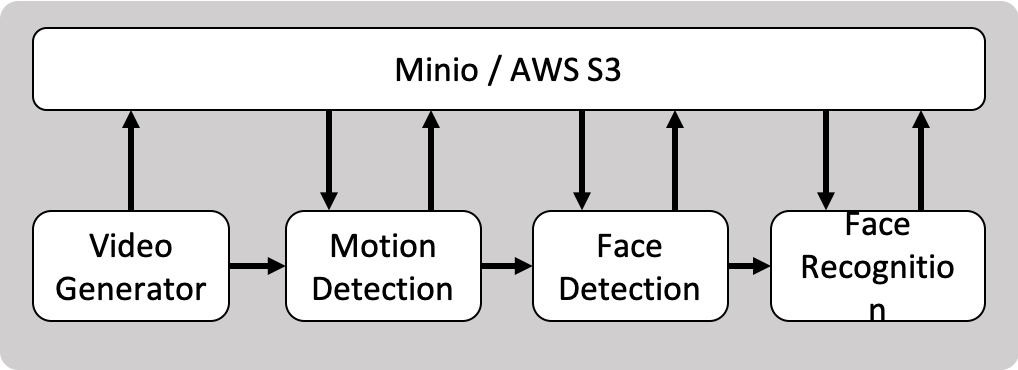}
	\caption{Video Analytics Workflow}
	\label{fig:ebench7}
	\centering
\end{figure}

\smallskip
\noindent
\textbf{Video Analytics Workflow.}
The video analytics workflow models workloads that have continuous data streams. It implements an end-to-end pipeline workflow from when a video stream is generated to when the final video analytics result is stored. Figure~\ref{fig:ebench7} shows the architecture of the workflow. 
A video stream generator continuously generates live videos and chunks them into groups of pictures (GoP) using FFmpeg~\cite{ffmpeg}. These small video chunks are the output of the first stage and are used by the next function called motion detection as input. At this step frames that contain motion are detected and sorted as images. The motion detection function uses OpenCV~\cite{opencv} to do inter-frame comparison. The generated images are fed to the face detection function. This function detects the faces in the images using Single Shot MultiBox Detector (SSD)~\cite{liu2016ssd} and filters out images containing faces as the input for the last function in the workflow, face recognition function. This function first uses Convolutional Neural Networks (CNN) to encode each detected face and then uses k-nearest neighbors (KNN) algorithm to classify the faces. The final output is the images that are marked with identities. The intermediate data is transferred using persistent storage.

Comparing the four stages, the first two stages generate most I/Os since they need to process every piece of the video. Face recognition has the highest CPU demand since it involves two steps of computation (neural network and KNN algorithm). In this implementation, the storage backends are Minio and AWS S3, which can be customized by users. Users can also customize the number of concurrent video streams generated by the first stage and the use of accelerators such as GPUs.

\begin{figure}[t]
	\centering
	\includegraphics[width=0.95\columnwidth]{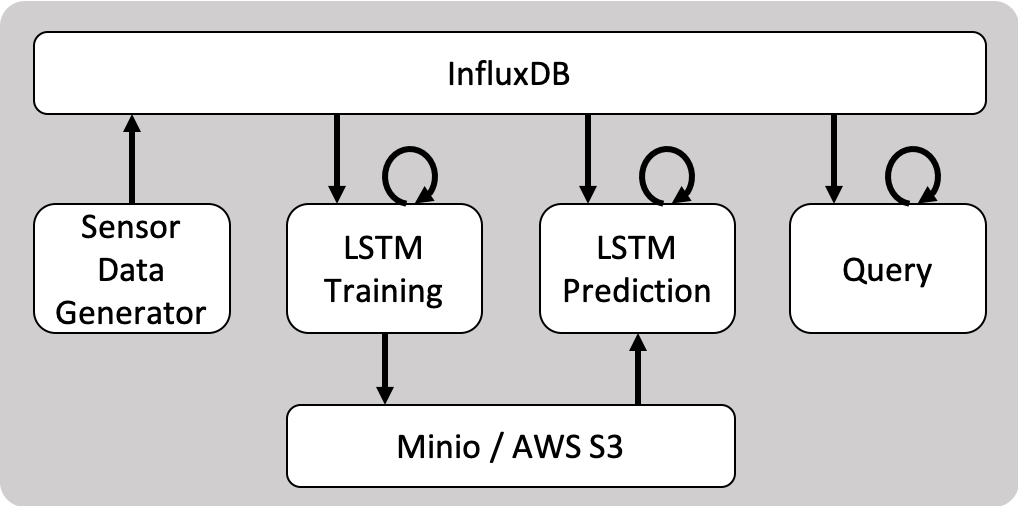}
	\caption{IoT Hub Workflow}
	\label{fig:ebench8}
	\centering
\end{figure}

\smallskip
\noindent
\textbf{IoT Hub Workflow.}
This workflow models workloads that generate vast amount of small and frequent data. Each IoT device sends diagnostic data frequently including the location (latitude, longitude, elevation) of the IoT device, the temperature of the environment, the moisture of the environment, the power consumption, and its health status. This workflow includes four components (Figure~\ref{fig:ebench8}). First, data generator continuously generates IoT devices' diagnostic data in a certain frequency to the database. We use MQTT for sensor data delivery from IoT devices to the database. Second, LSTM training function reads the latest data points from the database as input, trains a machine learning model using Long Short-Term Memory (LSTM)~\cite{hochreiter1997long} neural network and stores the model as output in the persistent storage. Third, LSTM prediction function reads history data from the database as the first input and the latest-trained machine learning model from the persistent storage as the second input, it then conducts prediction based on the history data and the LSTM-trained model. This stage prints out the predicted data as output. Forth, the query function generates a query on the IoT devices' diagnostic data in the database. The queries are from the TSBS benchmark~\cite{tsbs}. It reads latest database data as input and prints the query result as output.  

Except for the sensor data generator function which is a long-running job, the other three functions are set as cron workflows with different timers. Among the four jobs, LSTM training function is the most computation intensive as it conducts model training. Query function consumes the most memory as it loads a lot of data from database into the memory. Data generator generates most I/Os. In this implementation, we use InfluxDB and AWS Timestream as the time series database. Minio and AWS S3 are used as the storage backend for the machine learning model. Besides these user-customizable storage backends, users can also customize each cron job's timer, the input dataset size for LSTM training, LSTM prediction, and query, the number of concurrent requests triggered by each cron job, and the use of computing accelerators.

\vspace{-5pt}
\subsection{Reported Metrics}
\vspace{-5pt}
EdgeBench uses Prometheus~\cite{prometheus} to report system resource usages such as CPU, memory, I/O usage for the workflow and the running platform. To provide function-level metrics, EdgeBench also traces the function handler, storage backend load/store, and inter-function communication. These data can be used to provide each function's runtime latency, storage read/write latency, and the function communication latency, respectively. The data is automatically collected and stored in the database and get processed by the EdgeBench. 
\section{Evaluation}
\label{sec:eval}
\begin{table}[t]
	\centering
	\caption{Specifications of Computing Tiers.}
	\vspace{5pt}
	\small
	\tabcolsep=0.11cm
	\begin{tabular}{|c|c|c|c|}
		\hline
		&\textbf{Cloud tier}&\textbf{Edge tier}	&\textbf{IoT tier} \TBstrut\\    
		\hline
		   \multirow{3}{*}{\textbf{CPU}}  &Xeon  & Xeon E5& ARM	\Tstrut\\ 
		    & Platinum   &2630 V3  & Cortex - A72	\\ 
		            & 8175M  &  &	\Bstrut\\ 
		\hline	
		\textbf{RAM}& 128 GB &64 GB  &4 GB    \TBstrut\\    
		\hline
		\multirow{2}{*}{\textbf{Storage}}& 30 GB EBS& 400 GB  &64 GB   \Tstrut\\   
		          & volume& NVMe SSD &SD Card  \Bstrut \\    
		\hline
		\textbf{Operating }&Amazon  &Ubuntu   &Raspbian Stretch \Tstrut \\  
		         \textbf{System} &Linux 2 &18.04  & Lite 2020  \Bstrut \\ 
		\hline
		\textbf{Number of }&\multirow{2}{*}{6 - 10} &\multirow{2}{*}{9}  & \multirow{2}{*}{10}  \Tstrut\\   
		           \textbf{Nodes}& &  &   \Bstrut\\    
		\hline
	\end{tabular}
	\vspace{-10pt}
	\label{table:devices}
\end{table}

\begin{figure*}[t]
	\begin{subfigure}[b]{0.48\textwidth}
		\includegraphics[width=1.0\textwidth]{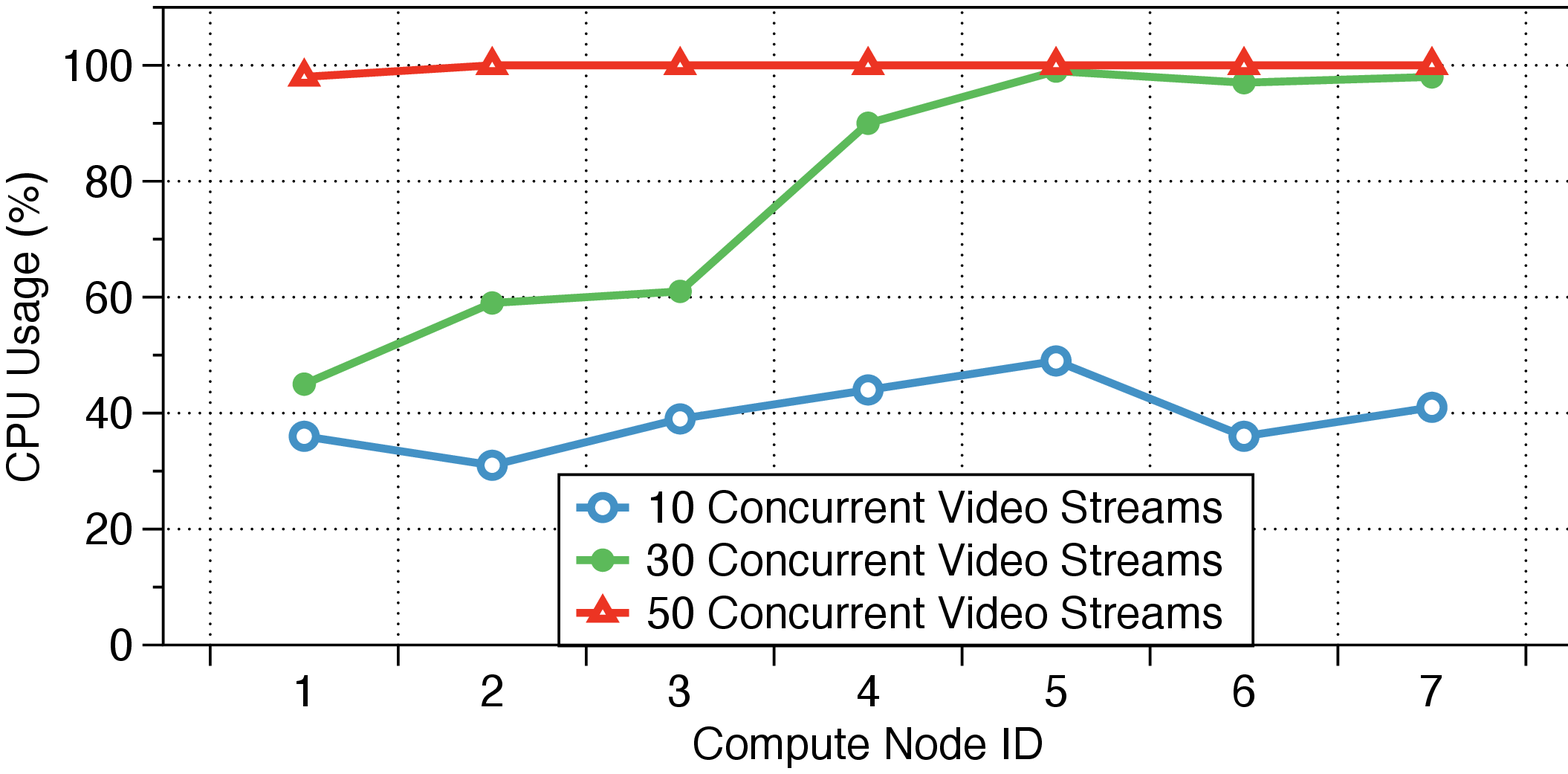}
		\caption{CPU Usage}
		\label{fig:1.1}
	\end{subfigure}
	\hspace{5pt}
	\begin{subfigure}[b]{0.48\textwidth}
		\includegraphics[width=1.0\textwidth]{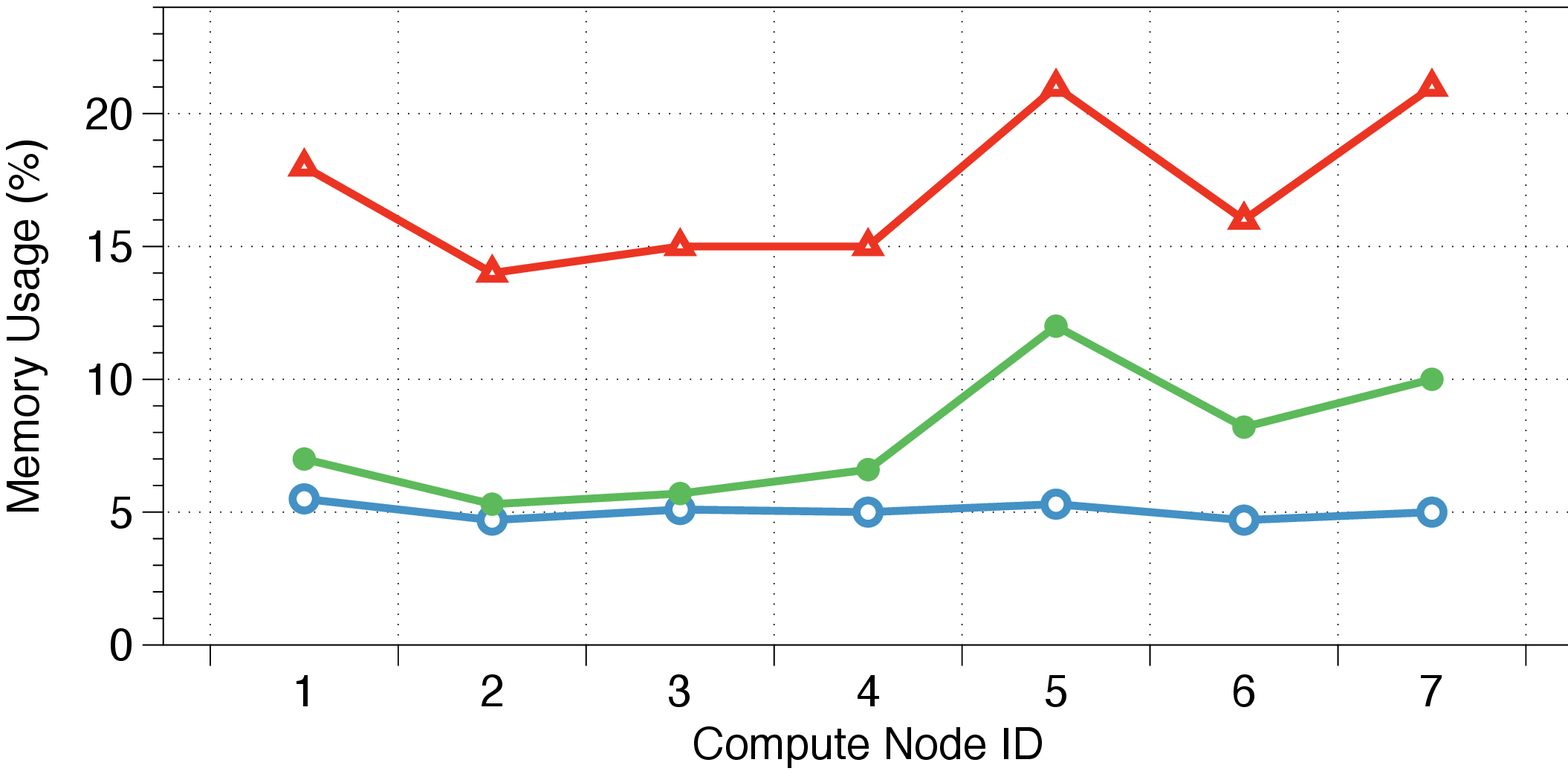}
		\caption{Memory Usage}
		\label{fig:1.2}
	\end{subfigure}
	\begin{subfigure}[b]{0.48\textwidth}
		\includegraphics[width=1.0\textwidth]{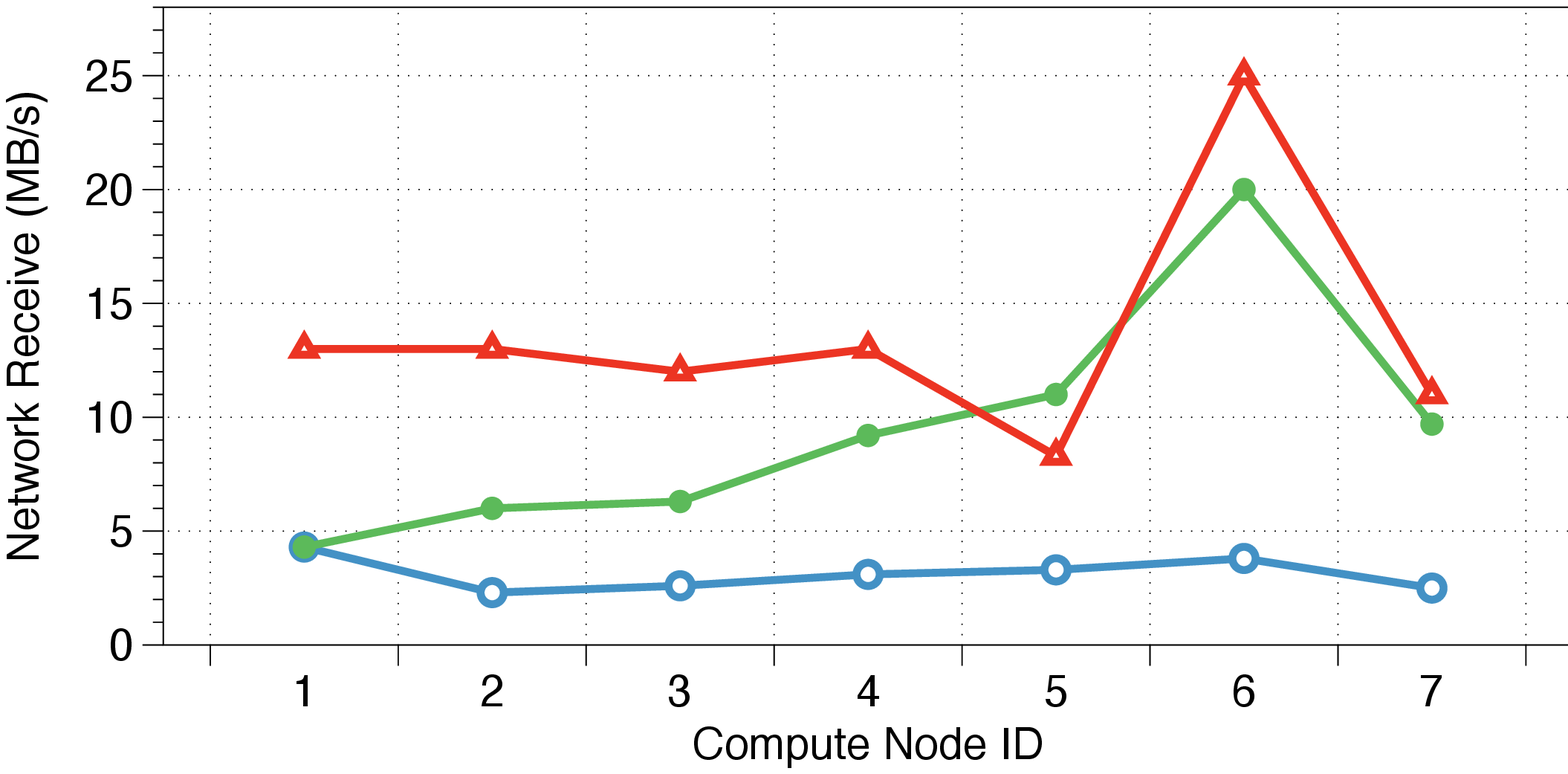}
		\caption{Network Receive Throughput}
		\label{fig:1.3}
	\end{subfigure}
	\hspace{10pt}
	\begin{subfigure}[b]{0.48\textwidth}
		\includegraphics[width=1.0\textwidth]{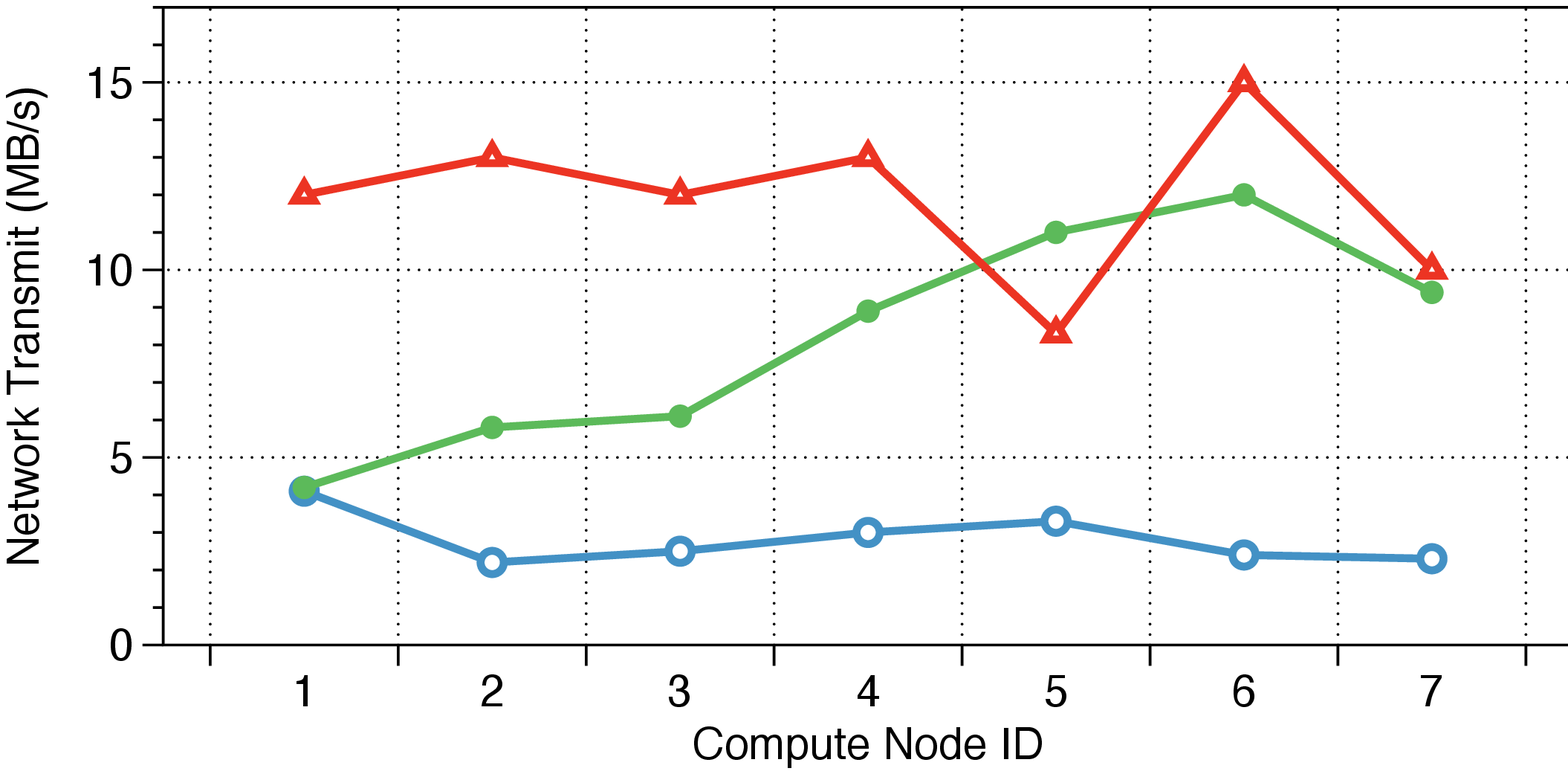}
		\caption{Network Transmit Throughput}
		\label{fig:1.4}
	\end{subfigure}
	\caption{Resource Usage of Each Node}
	\label{fig:1}
	\vspace{-13pt}
\end{figure*}

\begin{figure}[t]
	\hspace{10pt}
	\begin{subfigure}[b]{0.224\textwidth}
		\includegraphics[width=1.0\textwidth]{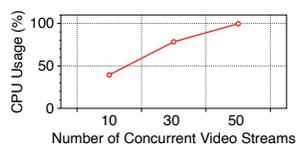}
		\caption{CPU Usage}
		\label{fig:2.1}
	\end{subfigure}
	\begin{subfigure}[b]{0.224\textwidth}
		\includegraphics[width=1.0\textwidth]{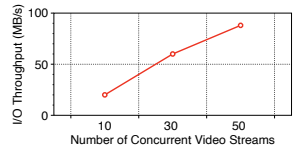}
		\caption{I/O Throughput}
		\label{fig:2.2}
	\end{subfigure}
	\caption{Video Analytics Workflow Resource Usage}
	\label{fig:2}
	\vspace{-13pt}
\end{figure}

We evaluated EdgeBench based on the two workflows and the prototype implemented on OpenFaaS. The evaluation setup involves three tiers: the IoT tier, the edge tier and the cloud tier. Each tier's specifications are shown in Table~\ref{table:devices}. The IoT tier is a cluster of ten Raspberry Pi 4 devices, which uses K3S~\cite{k3s} with OpenFaaS. The Edge tier is the nearby ASU cluster. The cluster consists of nine identical edge servers and uses Kubernetes with OpenFaaS. The Cloud tier uses the AWS Elastic Kubernetes Service (EKS)~\cite{eks}. The backend uses m5.8xlarge AWS EC2~\cite{ec2} instances as worker nodes. The cloud tier scales the number of worker nodes from 6 to 10 and uses Kubernetes with OpenFaaS. Of all the three tiers, one node serves as the master node that manages all the services and one node serves as the data node that hosts persistent storage and database. All the other nodes are compute nodes. 

A 10FPS, 1920 x 1080 video file is used as the video source for reproducible results. For the IoT hub workflow, each IoT device is set to send sensor data every second. LSTM training is set to execute every half an hour. The training process is based on the data generated within the last 30 minutes. LSTM prediction function is set to be invoked every 5 seconds. It conducts prediction based on the data generated within the last 30 seconds. The query function is set to be invoked every 3 seconds, every time it randomly picks one query from the query pool that has 12 queries and executes.  

InfluxDB and Minio are deployed on Kubernetes, both use data node's local storage device as the persistent storage. During the experiments, we set OpenFaaS to auto scale the functions from 25 to 100 when load increases at a factor ratio of 25\% and it automatically scales down the number of function containers when the load drops. We ran each experiment for 5 times and picked the round with the most representative data to show the results. 

\subsection{Edge Tier Evaluation}
The first set of experiments run only on the edge tier to evaluate both the edge system and the edge workloads. We vary the number of concurrent requests issued to each workflow to evaluate the resource usage and performance of the workflow under different workload stress.

\smallskip
\noindent
\textbf{Video Analytics Workflow.} We ran video analytics workflow with three workload settings: 10, 30, and 50 concurrent video streams. For each load setting, the workflow runs for 30 minutes.

Figure~\ref{fig:1} shows the average CPU usage, memory usage, and network receive/transmit throughput of each compute node under different load stress. For different compute nodes, the workload is not evenly distributed. With 30 concurrent video streams, node 5 has the highest CPU utilization which is 99\% while node 1 has the lowest CPU utilization which is only 45\%. Similar fluctuation happened in memory and network usage. For example, with 50 concurrent video streams, node 5 has used 21\% of memory capacity while node 2 only used 14\% of memory capacity. 
It can also be observed that the resource (CPU, memory, and network) usage  is not growing proportional to the increased load in the system. 
When the load grows from 10 concurrent video streams to 30 concurrent video streams, node 1's CPU utilization only grows by 20\% while node 6's CPU utilization is 1.7 times higher. This uneven distribution of load can easily cause some compute nodes becoming the performance bottlenecks while some compute nodes are severely under utilized. This can further slow down the edge workload's performance and cause high latency.

\begin{figure}[t]
	\centering
	\includegraphics[width=0.42\textwidth]{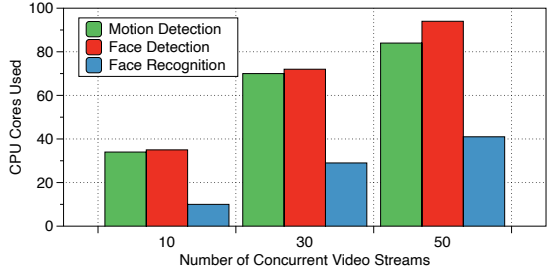}
	\caption{CPU Usage of Each Workflow Stage}
	\label{fig:3}
	\vspace{-6pt}
	\centering
\end{figure}

\begin{figure}[t]
	\centering
	\includegraphics[width=0.43\textwidth]{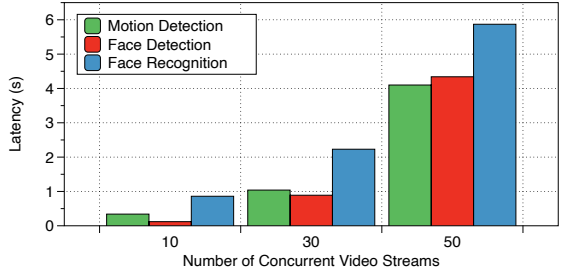}
	\caption{Latency of Each Workflow Stage}
	\label{fig:4}
	\vspace{-6pt}
	\centering
\end{figure}

\begin{figure}[t]
	\begin{subfigure}[b]{0.42\textwidth}
		\includegraphics[width=1.0\textwidth]{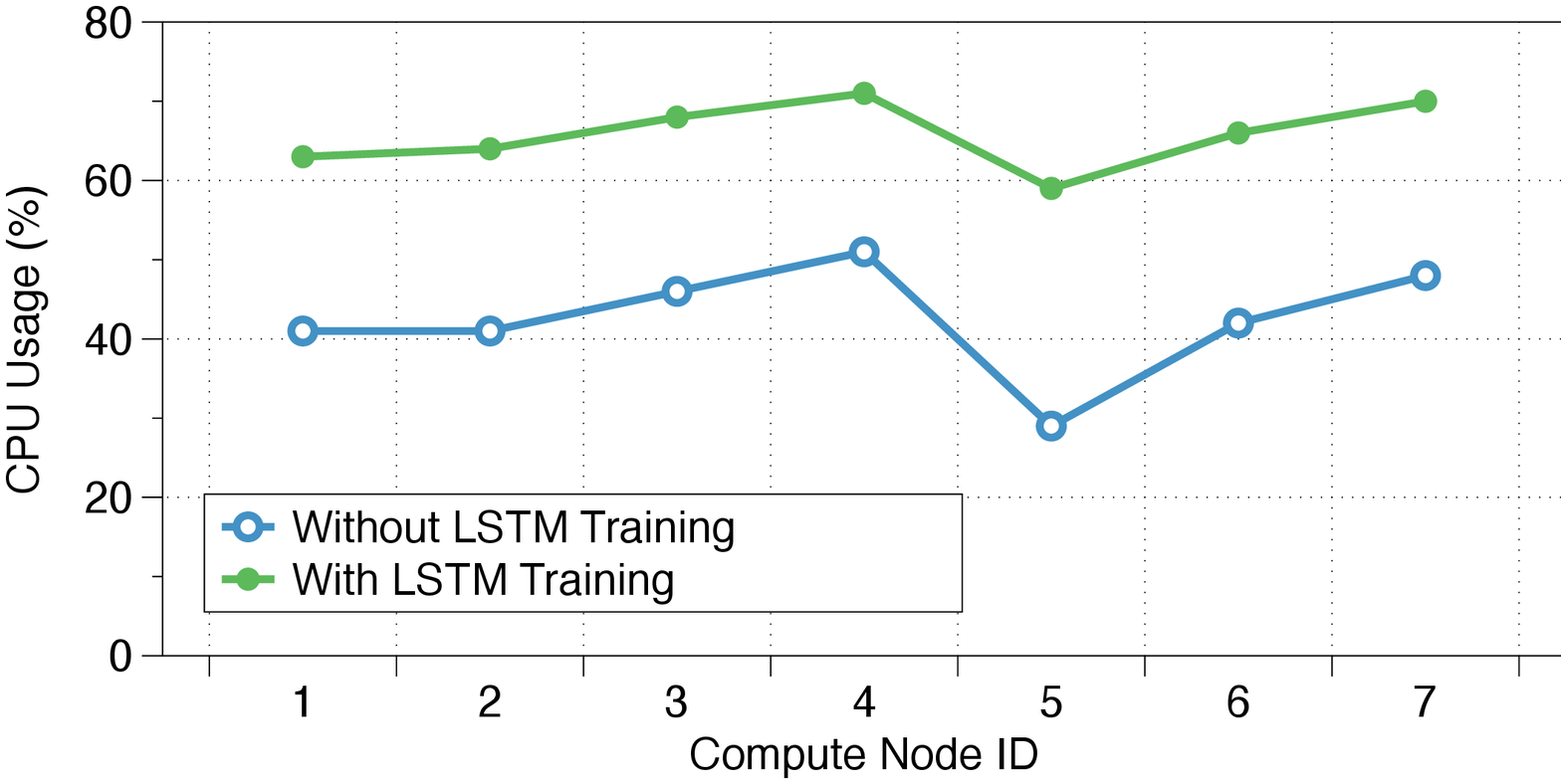}
		\caption{CPU Usage}
		\label{fig:5.1}
	\end{subfigure}
	\hspace{-5pt}
	\begin{subfigure}[b]{0.42\textwidth}
		\includegraphics[width=1.0\textwidth]{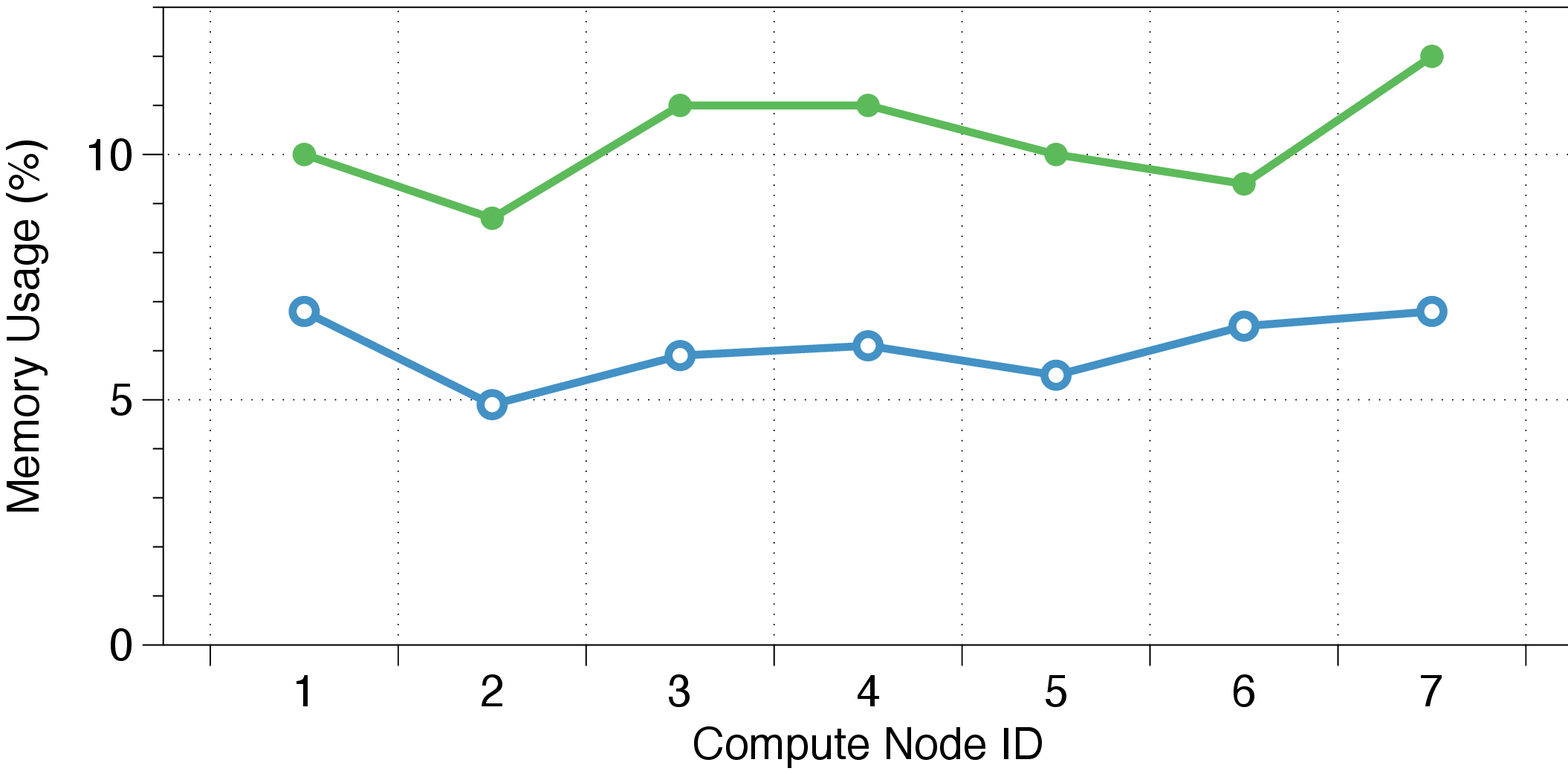}
		\caption{I/O Throughput}
		\label{fig:5.2}
	\end{subfigure}
	\caption{Resource Usage of Each Node}
	\label{fig:5}
	\vspace{-13pt}
\end{figure}

\begin{figure}[t]
	\centering
	\includegraphics[width=0.45\textwidth]{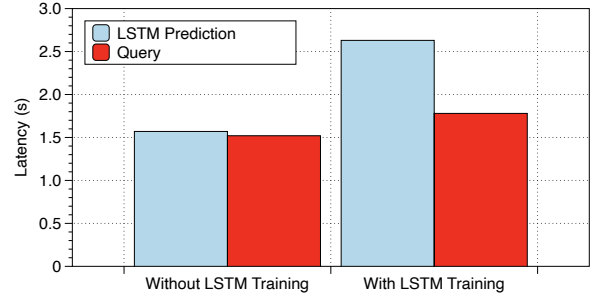}
	\caption{Latency With and Without LSTM Training}
	\label{fig:6}
	\vspace{-10pt}
	\centering
\end{figure}

\begin{figure*}[t]
	\hspace{-9pt}
	\begin{minipage}{0.34\textwidth}
		\begin{subfigure}[b]{\textwidth}
			\includegraphics[width=\textwidth]{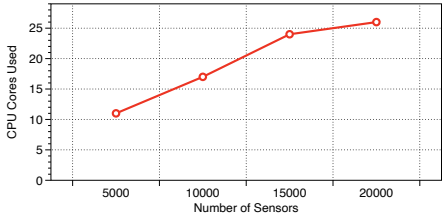}
			\caption{CPU Usage}
			\label{fig:7.1}
		\end{subfigure}
	\end{minipage}
	\hspace{2pt}
	\begin{minipage}{0.34\textwidth}
		\begin{subfigure}[b]{\textwidth}
			\includegraphics[width=\textwidth]{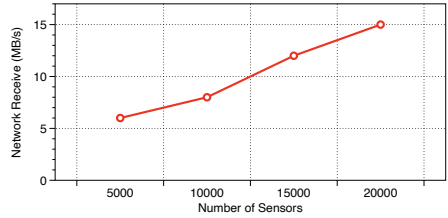}
			\caption{Network Receive Throughput}
			\label{fig:7.2}
		\end{subfigure}
	\end{minipage}
	\hspace{2pt}
	\begin{minipage}{0.34\textwidth}
		\begin{subfigure}[b]{\textwidth}
			\includegraphics[width=\textwidth]{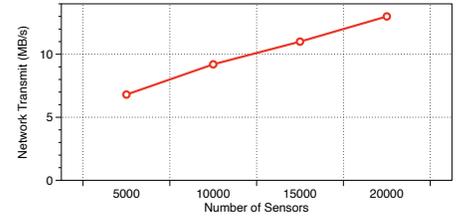}
			\caption{Network Transmit Throughput}
			\label{fig:7.3}
		\end{subfigure}
	\end{minipage}
	\caption{Resource Usage of MQTT}
	\label{fig:7}
	\vspace{-6pt}
\end{figure*}

Figure~\ref{fig:2} shows the average CPU utilization and I/O throughput of the video analytics workflow for different amount of concurrent video streams. From the figures, we can observe that CPU has been used up to 99.7\% when 50 concurrent video streams are processed. I/O throughput, however,  grows linearly and the bandwidth of the NVMe SSD has not been fully utilized. This observation shows that the video analytics workflow is more CPU bounded rather than I/O bounded.

We further investigated this workflow by evaluating the results of different stages: motion detection, face detection, and face recognition.
Figure~\ref{fig:3} illustrates the CPU utilization of each stage under different load stress. The number of CPU cores used by each stage grows with the number of concurrent video streams issued. Increasing the number of concurrent streams from 30 to 50, the growth rate drops. Because, the CPU starts to become the bottleneck with 50 concurrent video streams. Although face recognition is the most CPU intensive stage, it is not using the most amount of CPU cores in the workflow. This is because the previous two stages have filtered out images which contained no faces. As a result, compared to the previous two stages, the face recognition is not triggered frequently. 
In our video streams, approximately 18\% of the video frames go to the face recognition stage.

Figure~\ref{fig:4} shows the 95th percentile latency for each compute stage to finish one execution. Face recognition is the most computation intensive stage and takes the longest time to finish one execution since it involves two steps of CNN encoding and KNN classification. Motion detection and face detection take roughly the same amount of time. Motion detection runs a bit longer than face detection since it involves expensive video decoding. With increased workloads, the latency has increased for all the three stages. However, the latency increases more when the load increases from 30 to 50 concurrent video streams compared to that from 10 to 30 concurrent video streams. This shows that when the workload is intensive, the impact of container interference on the performance is more severe.

\smallskip
\noindent
\textbf{IoT Hub Workflow.} 
Contrary to the video analytics workflow, the IoT hub workflow is not a pipeline and each stage runs separately. In the experiment, we stress each stage with different workload settings. We first show EdgeBench as a macrobenchmark with the results of running the four stages altogether and then show EdgeBench as a microbenchmark where individual results of running each stage separately are presented. 

Since LSTM training is the most CPU intensive stage among the four, we would like to compare the resource usage and performance of the workflow with and without LSTM training running to see the impact of LSTM training on other jobs. To run the four jobs together, we first start sensor data generator, LSTM prediction, and query. LSTM training is triggered 20 minutes after the other three stages start. LSTM training takes more than an hour to finish one training process, we stop the four stages 40 minutes after the initial start. In the experiment, we set 15000 sensors sending sensor data concurrently in the sensor data generator. LSTM training issues 20 concurrent training requests, LSTM prediction issues 80 concurrent prediction requests, and query issues 20 concurrent queries at a time. We do not vary the stress of workload here but show workload stress results later when each stage runs separately.

Figure~\ref{fig:5} shows the CPU and memory usage of each compute node with and without LSTM training. As what we observed from the video analytics workflow, the load distribution is also imbalanced here. When LSTM training is not running, the highest CPU usage is 51\% for node 4 while node 5 only uses 29\% of the CPU. Another observation is that after LSTM training starts, both the CPU usage and memory usage have increased a lot. On average, the CPU usage increases by 54\% and the memory usage increases by 69\% with LSTM training running compared to LSTM training not running.

Besides the resource usage, we also observe the 95th percentile latency changes for the other workflow jobs with and without LSTM training. Figure~\ref{fig:6} compares LSTM prediction and query's latency changes when LSTM training runs or not. Both jobs' latency get prolonged. For LSTM prediction, it was affected more severely, where the latency increases from 1.57s to 2.63s by 67\%. 

We then show the results of running each job separately. In this experiment, each job is stressed with different number of concurrent requests to show that job's resource usage and performance. We vary the number of sensors from 5000 to 20000 in sensor data generator. For LSTM training, we issue concurrent requests from 10 to 25 increased by 5 requests. For LSTM prediction, the workload varies from 60 concurrent requests to 100 concurrent requests increased by 20 requests. For query, the workloads changes from 10 concurrent requests to 30 concurrent requests increased by 10 requests.

\begin{figure}[t]
	\begin{subfigure}[b]{0.224\textwidth}
		\includegraphics[width=1.0\textwidth]{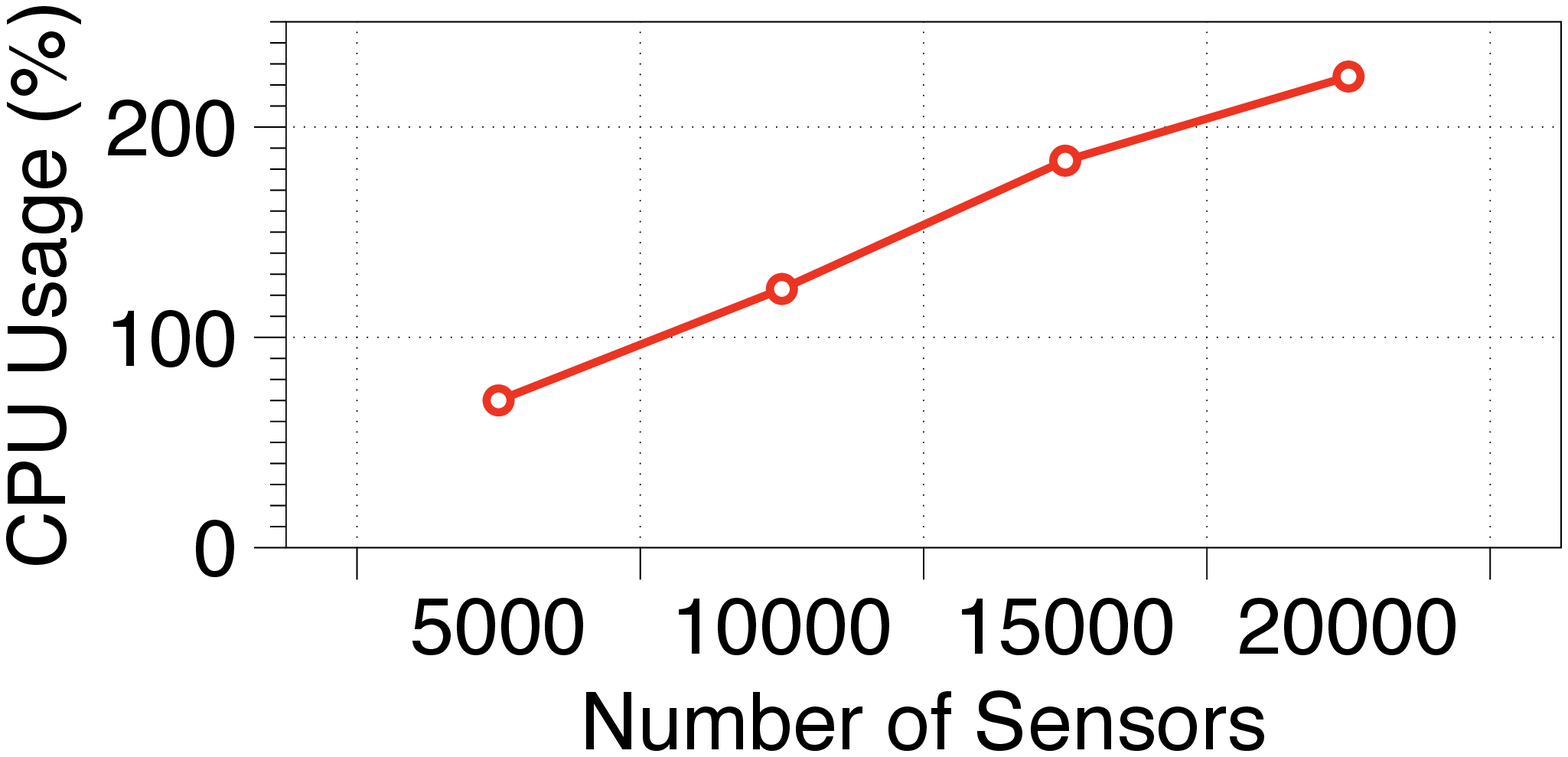}
		\caption{CPU Usage}
		\label{fig:8.1}
	\end{subfigure}
	\hspace{6pt}
	\begin{subfigure}[b]{0.224\textwidth}
		\includegraphics[width=1.0\textwidth]{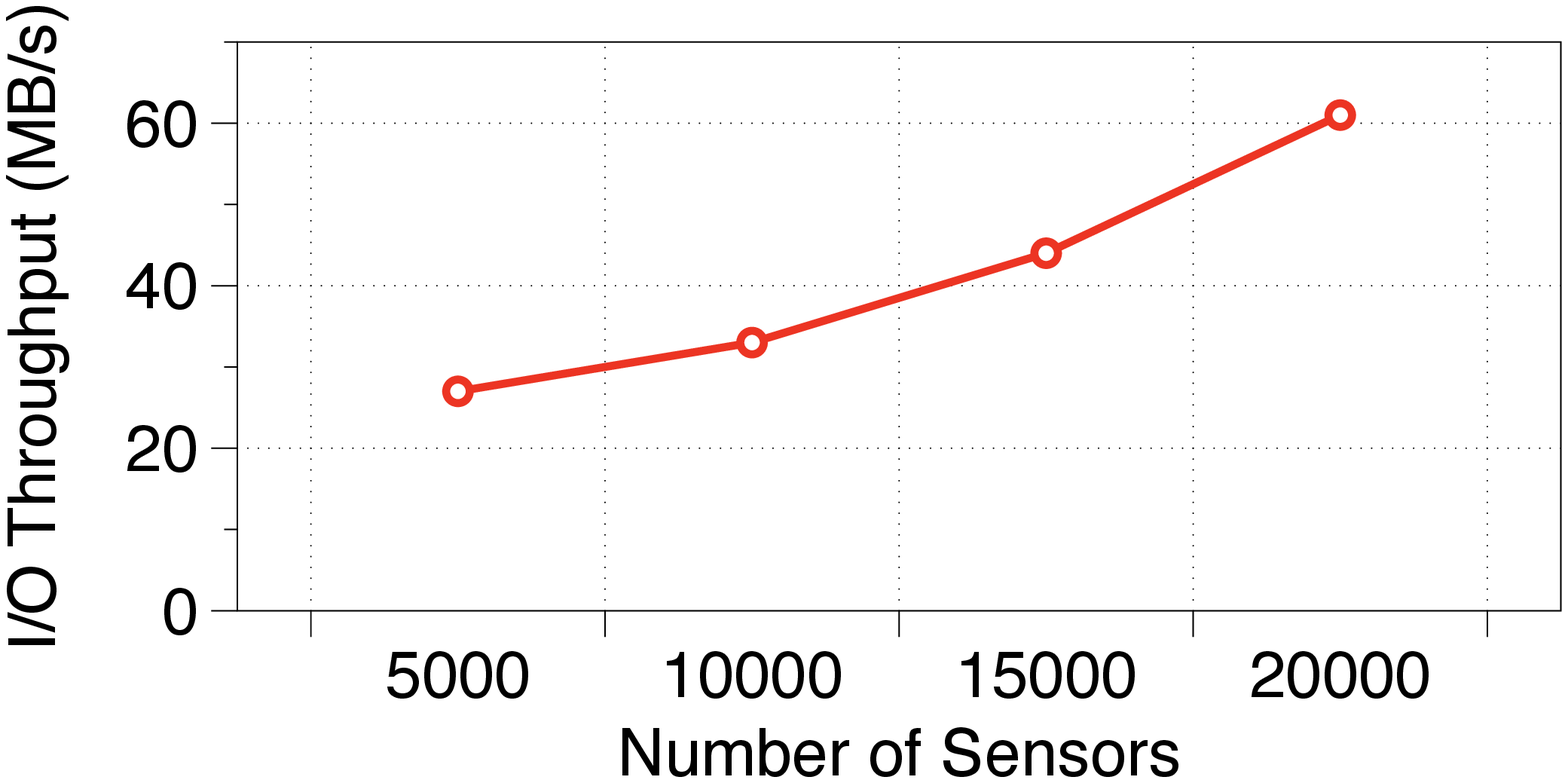}
		\caption{I/O Throughput}
		\label{fig:8.2}
	\end{subfigure}
	\caption{Resource Usage of InfluxDB}
	\label{fig:8}
	\vspace{-13pt}
\end{figure}

Figure~\ref{fig:7} and Figure~\ref{fig:8} show the experiment results from sensor data generator. Since sensor data generator mainly uses MQTT for data delivery and InfluxDB for data storage, we show the resource usage of these two services. Figure~\ref{fig:7.1} shows the number of CPU cores used by MQTT under different workloads. When workload changes from 15000 sensors to 20000 sensors, the CPU usage is not increased linearly which shows that CPU has become the bottleneck with 20000 sensors. In Figure~\ref{fig:7.2} and Figure~\ref{fig:7.3}, the network receive/transmit throughput increases linearly with the workload which illustrates network is not stressed with this many concurrent sensors. Figure~\ref{fig:8} shows the InfluxDB CPU usage and I/O throughput. Both CPU usage and I/O throughput increases linearly with the workload increase and database insertion is not aggressively stressing the database.   

\begin{figure*}[t]
	\hspace{10pt}
	\begin{subfigure}[b]{0.45\textwidth}
		\includegraphics[width=1.0\textwidth]{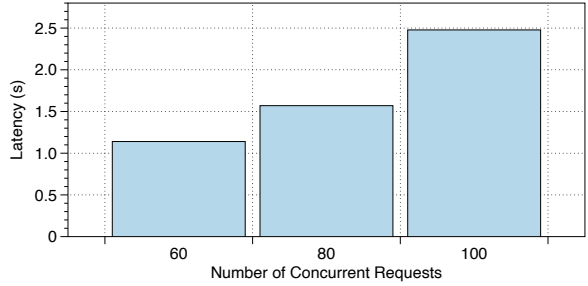}
		\caption{LSTM Prediction}
		\label{fig:9.1}
	\end{subfigure}
	\hspace{20pt}
	\begin{subfigure}[b]{0.45\textwidth}
		\includegraphics[width=1.0\textwidth]{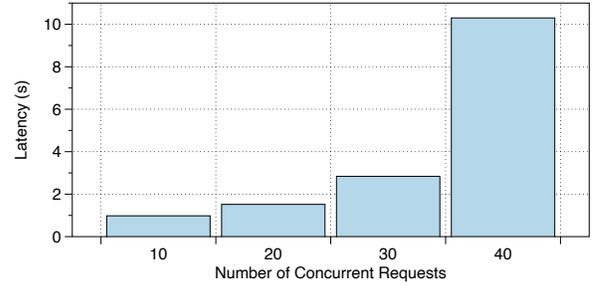}
		\caption{Query}
		\label{fig:9.2}
	\end{subfigure}
	\caption{Latency Changes with Different Stress of Loads}
	\label{fig:9}
	\vspace{-10pt}
\end{figure*}

For LSTM training, LSTM prediction and query, the CPU and memory usage for each stage increases linearly with the workload increases and none of them have caused CPU or memory bottleneck. Figure~\ref{fig:9} shows the 95th percentile latency increase with incremental stress of workloads for LSTM prediction and query. Similar as what we observed in the video analytics workflow, the latency increases more when the workload gets heavier.

\subsection{Cross Tier Evaluation}
The second set of experiments run across the IoT tier, the edge tier, and the cloud tier to evaluate which distribution setting works the best for each workflow. We show the 95th percentile latency of each stage for both workflows and the end-to-end latency of the video analytics workflow. For each workflow, we have three distribution settings: IoT tier and edge tier, IoT tier and cloud tier, and three tiers combined.

\begin{table}[t]
	\centering
	\caption{Cross Tier Distribution of Video Analytics Workflow}
	\vspace{5pt}
	\small
	\tabcolsep=0.11cm
	\begin{tabular}{|c|c|c|c|c|}
		\hline
		Settings&\textit{Video}&\textit{Motion}	&\textit{Face}&\textit{Face} \Tstrut\\  
		&\textit{Generator}&\textit{Detection}	&\textit{Detection}&\textit{Recognition} \Bstrut\\  
		\hline
		\textit{IoT and edge}&IoT  & IoT & Edge	& Edge \TBstrut\\ 
		\hline	
		\textit{IoT and cloud}& IoT &IoT &Cloud& Cloud   \TBstrut\\    
		\hline
		\textit{three tiers}& IoT &IoT &Edge& Cloud   \TBstrut \\   
		\hline
	\end{tabular}
	\label{table:distribution1}
\end{table}

\begin{figure}[t]
	\centering
	\includegraphics[width=0.45\textwidth]{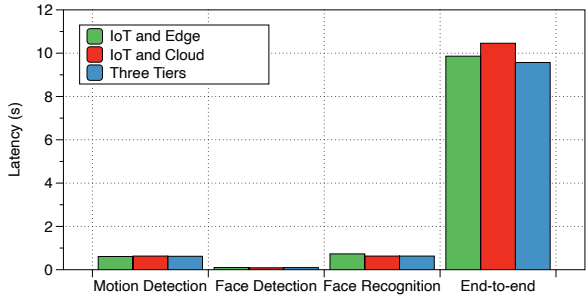}
	\caption{Latency of Video Analytics Workflow Across Tiers}
	\label{fig:11}
	\vspace{-10pt}
	\centering
\end{figure}

\begin{figure*}[t]
	\hspace{-9pt}
	\begin{minipage}{0.34\textwidth}
		\begin{subfigure}[b]{\textwidth}
			\includegraphics[width=\textwidth]{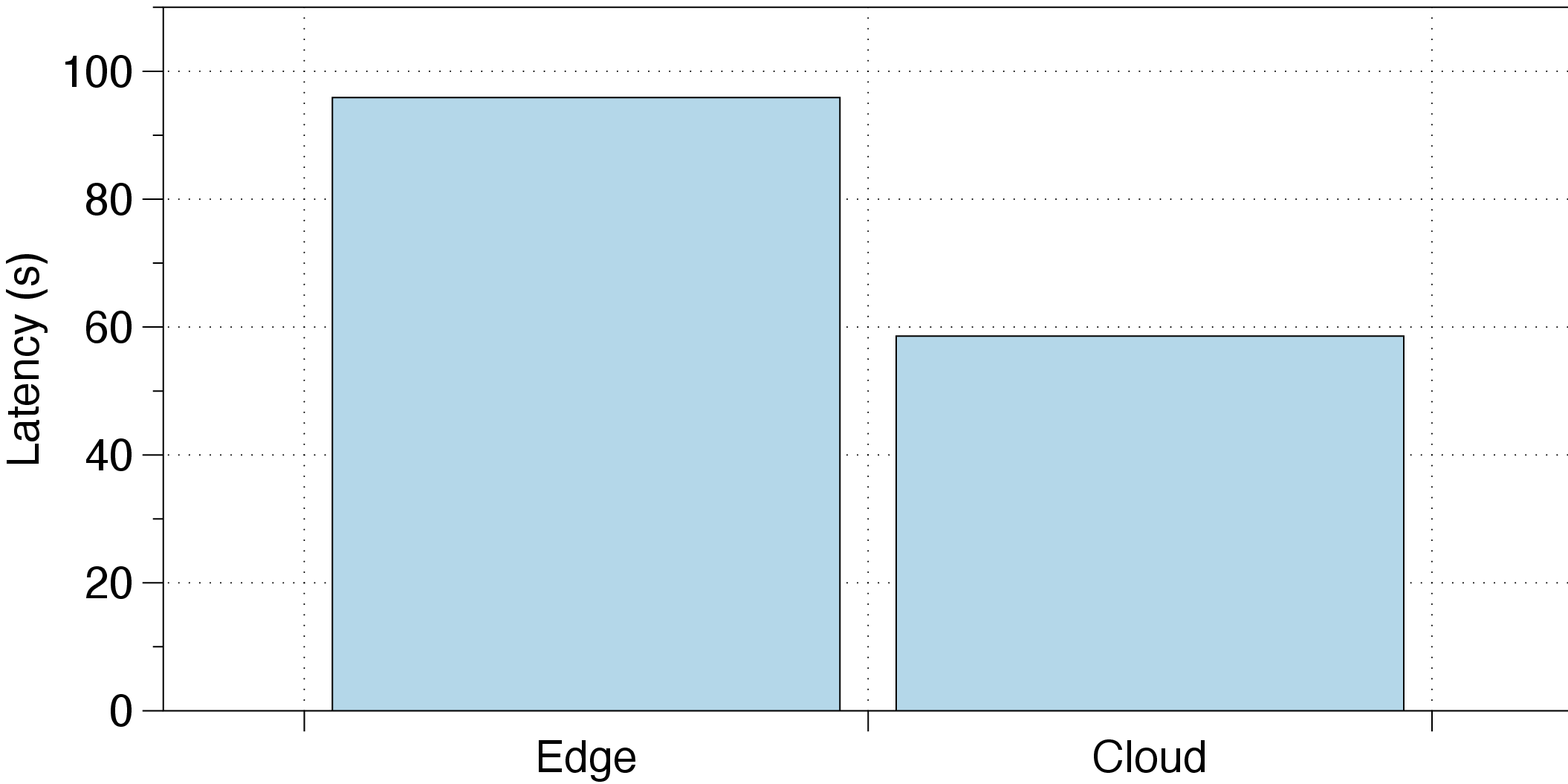}
			\caption{LSTM Training}
			\label{fig:10.3}
		\end{subfigure}
	\end{minipage}
	\hspace{2pt}
	\begin{minipage}{0.34\textwidth}
		\begin{subfigure}[b]{\textwidth}
			\includegraphics[width=\textwidth]{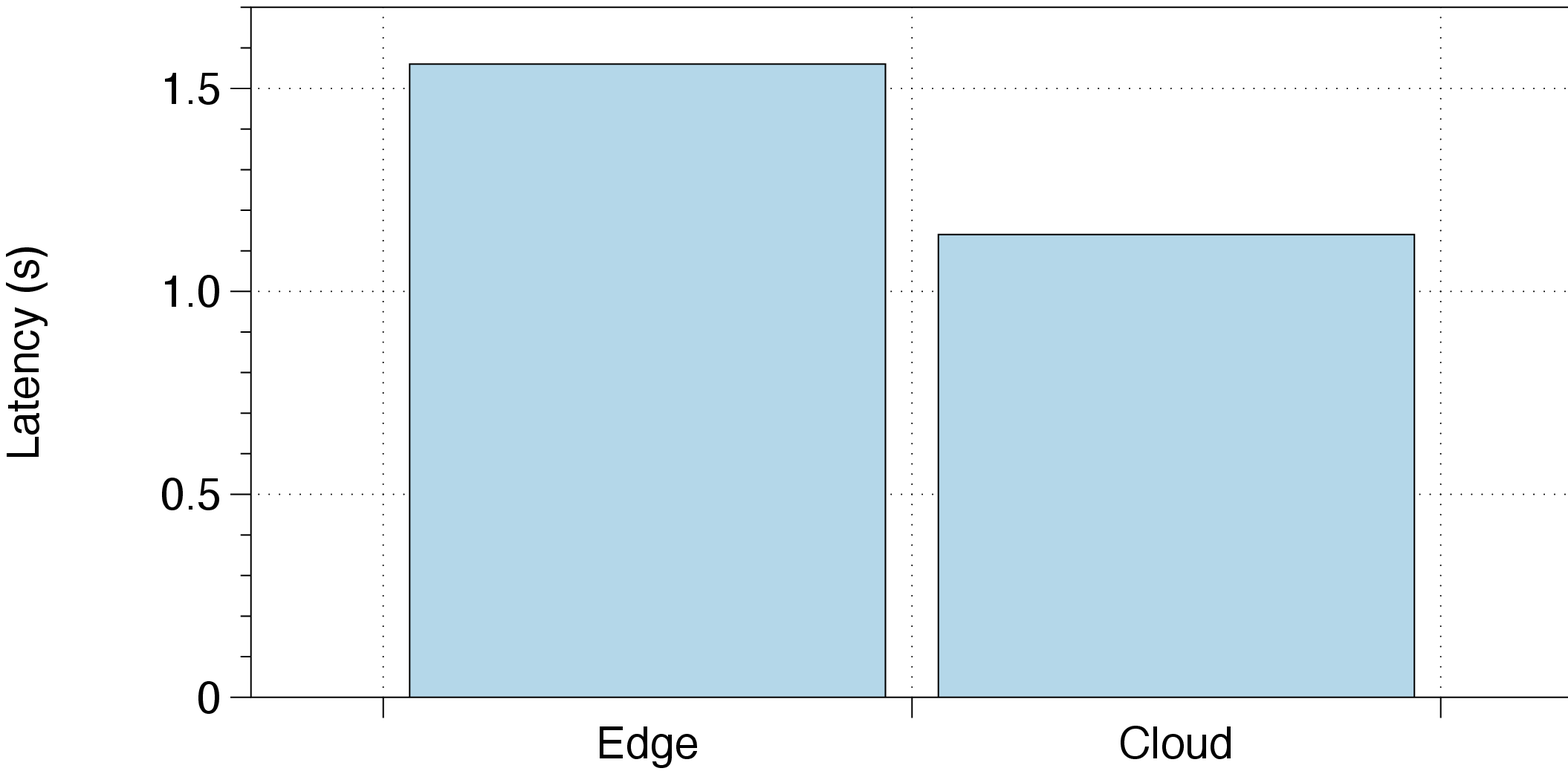}
			\caption{LSTM Prediction}
			\label{fig:10.2}
		\end{subfigure}
	\end{minipage}
	\hspace{2pt}
	\begin{minipage}{0.34\textwidth}
		\begin{subfigure}[b]{\textwidth}
			\includegraphics[width=\textwidth]{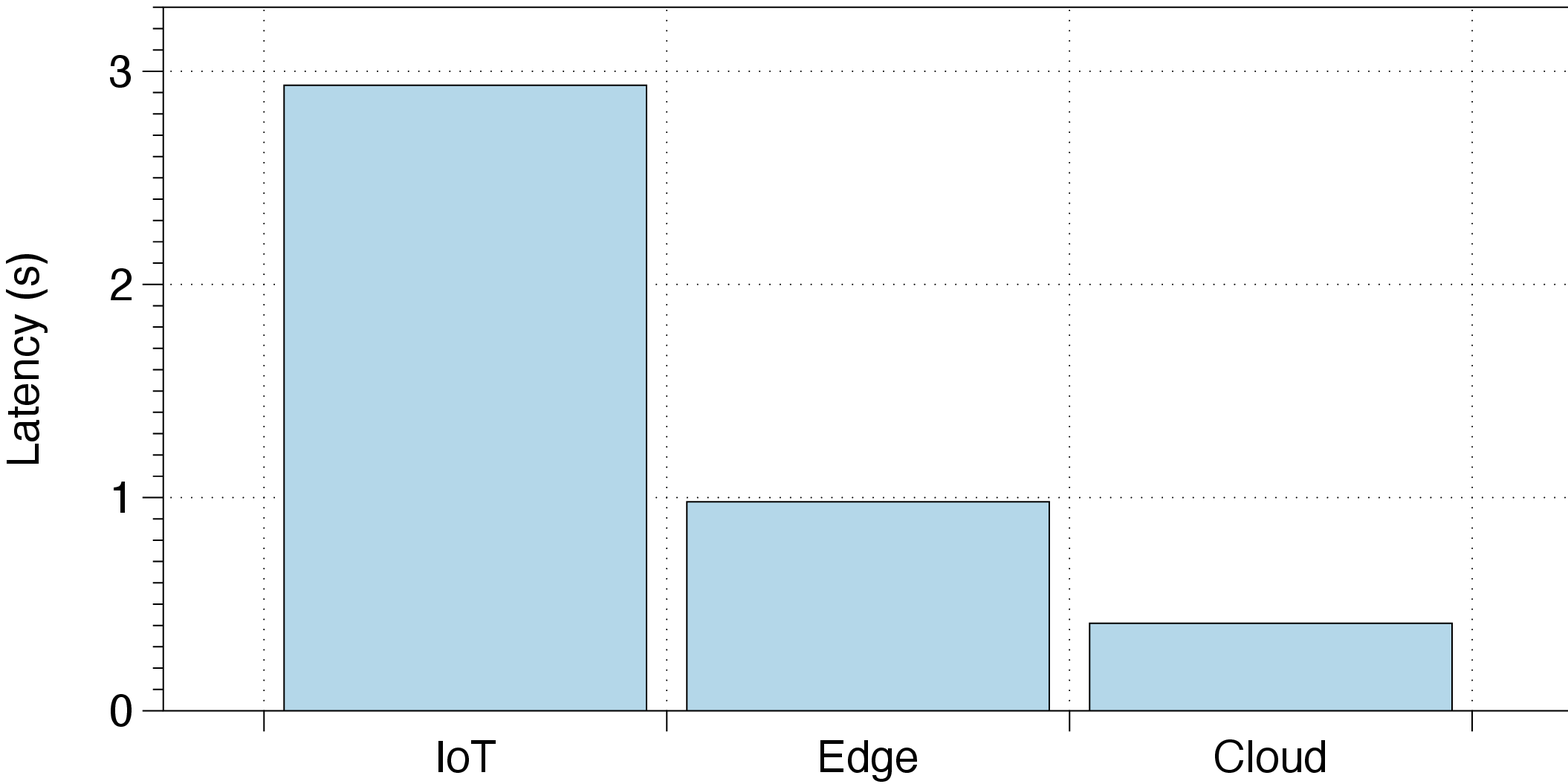}
			\caption{Query}
			\label{fig:10.1}
		\end{subfigure}
	\end{minipage}
	\caption{Latency of IoT Hub Workflow Across Tiers}
	\label{fig:10}
	\vspace{-6pt}
\end{figure*}

\smallskip
\noindent
\textbf{Video Analytics Workflow.} The cross tier distribution of the video analytics workflow under each setting is shown in Table~\ref{table:distribution1}. The first two stages are distributed on the IoT tier to save the network latency. The computation intensive face detection and face recognition are distributed on the edge tier or cloud tier. For the IoT and edge tiers, we have Minio deployed on that tier's devices. For the cloud tier, we use AWS S3 for storage. For each setting, we issue 10 concurrent video streams for 30 minutes considering IoT devices capacity.

Figure~\ref{fig:11} shows the latency result. We present the latency of each stage and the end-to-end latency of each request (Video generator result is not reported here since it is a long-running job). Compare \textit{IoT and edge} to \textit{IoT and cloud}, each stage has a smaller latency when execute on the cloud tier. However, due to the high network latency when communicate across stages, the overall end-to-end latency of \textit{IoT and cloud} outweighs that of the \textit{IoT and edge} by 6\%. \textit{Three tiers} takes both the advantages of the computation power of the cloud tier and the low network latency of the edge tier and has the lowest end-to-end latency. The video analytics workflow should be applied across the three tiers to utilize the benefits from different resources.

\smallskip
\noindent
\textbf{IoT Hub Workflow.} 
The IoT hub workflow has four separate running jobs, so we show the latency of running each job on the IoT tier, the edge tier and the cloud tier here. Sensor data generator is a long-running job where latency is not reported. During the experiments, we run sensor data generator all the time to generate updated data to the database. LSTM Training usually takes hours to run, to show the latency, we reduce the number of epochs of the training process to reduce the training time. Since LSTM training and LSTM prediction are computation intensive and not suitable to run on IoT devices, we only show the edge tier and cloud tier latency for the two jobs. 
To run jobs on the IoT tier and the edge tier, we deploy InfluxDB on the edge tier. For the cloud tier, we use AWS Timestream database for data storage. The machine learning model used by LSTM training and LSTM prediction is stored to Minio on the edge tier. For the cloud tier, we use AWS S3 for model storage. 

Every time the cron jobs trigger, we issue 10 concurrent requests for LSTM training, 60 concurrent requests for LSTM prediction, and 10 concurrent requests for query, respectively. Each experiment runs for 30 minutes.

Figure~\ref{fig:10} shows the latency of LSTM training, LSTM prediction and query when running on the IoT tier, the edge tier, and the cloud tier. For query, the IoT tier takes the longest time to finish due to the resource limitation of IoT devices. Compared to the cloud tier latency, it has 6 times overhead. For all the three jobs, the edge tier takes longer time to finish each request than the cloud tier. LSTM training, as the most computation intensive job, has the largest overhead of 64\% by running on the edge tier compared to the cloud tier. For the IoT hub workflow, users should distribute the computation intensive job LSTM training on the cloud tier to improve its performance and reduce its interference with other jobs.

\eat{\begin{table}[h]
	\centering
	\caption{Cross Tier Distribution of IoT Hub Workflow}
	\vspace{5pt}
	\small
	\tabcolsep=0.11cm
	\begin{tabular}{|c|c|c|c|c|}
		\hline
		Settings&\textit{Sensor Data}&\textit{Query}	&\textit{LSTM}&\textit{LSTM} \Tstrut\\  
		&\textit{Generator}&\textit{}	&\textit{Prediction}&\textit{Training} \Bstrut\\  
		\hline
		\textit{IoT and edge}&IoT  & IoT & Edge	& Edge \TBstrut\\ 
		\hline	
		\textit{IoT and cloud}& IoT &IoT &Cloud& Cloud   \TBstrut\\    
		\hline
		\textit{three tiers}& IoT &IoT &Edge& Cloud    \TBstrut\\   
		\hline
	\end{tabular}

	\label{table:distribution2}
\end{table}}

\section{Related Works}
\label{sec:related}
\vspace{-5pt}

Edge workloads have attracted lots of attention and there are multiple edge benchmarks developed for generating edge workloads and evaluating edge applications performance~\cite{wang2018cavbench, mcchesney2019defog, gan2019open, sridhar2017evaluating}.

CAVBench~\cite{wang2018cavbench} is a benchmark suite for connected and autonomous vehicles. In their work, they provided several CAV workloads. DeFog~\cite{mcchesney2019defog} is a benchmark with six edge applications. Their applications cover more other areas of edge applications besides CAV applications. Both works provide workloads that are single components in an end-to-end edge workflow pipeline. The benchmarks lack representativeness of end-to-end complete edge workloads. 

DeathStarBench~\cite{gan2019open}, on the other hand, provides a benchmark suite that covers a complete edge application pipeline (the drone swarm application). Related work~\cite{sridhar2017evaluating} evaluates a voice interaction pipeline that provides an end-to-end edge application for voice interaction. However, the edge applications are predefined workflows and users are not allowed to customize each stage, data storage backends, and computing tiers like EdgeBench. 

\section{Conclusion}
\label{sec:con}
\vspace{-5pt}
We have presented EdgeBench, a workflow-based benchmark for edge computing. EdgeBench is customizable. It provides the user-defined workflow orchestration. Users can choose different data transfer and storage backends for different components of the workflow. Each individual function can be allocated to different computing tiers. EdgeBench is representative. The benchmark implements two representative edge workflow, the video analytics workflow and the IoT hub workflow. EdgeBench can be used as both micro and macro benchmarks with reported metrics both at function-level and workflow-level. We encourage users to use EdgeBench to study both the edge systems and edge workloads. In the future, we plan to extend EdgeBench to include more predefined workflow logics and storage backends as default options. 



{\bibliographystyle{ieeetr}
	\bibliography{main}}

\eat{
This demo file is intended to serve as a ``starter file''
for IEEE Computer Society conference papers produced under \LaTeX\ using
IEEEtran.cls version 1.8b and later.
I wish you the best of success.

\hfill mds
 
\hfill August 26, 2015

\subsection{Subsection Heading Here}
Subsection text here.

\subsubsection{Subsubsection Heading Here}
Subsubsection text here.}

\eat{}

\end{document}